\pdfoutput=1

\documentclass[11pt,twoside,a4paper,cmspaper,final,collab]{cms-tdr}
\def\svnVersion{257104}\def\cmsCernNo{2013-037}\def\cmsCernDate{\today}\def\cmsMessage{Published in Physical Review D as \href{http://dx.doi.org/10.1103/PhysRevD.90.032005}{\doi{10.1103/PhysRevD.90.032005.}}}

\begin{document}\cmsNoteHeader{EXO-12-051}

\hyphenation{had-ron-i-za-tion}
\hyphenation{cal-or-i-me-ter}
\hyphenation{de-vices}
\RCS$Revision: 253096 $
\RCS$HeadURL: svn+ssh://svn.cern.ch/reps/tdr2/papers/EXO-12-051/trunk/EXO-12-051.tex $
\RCS$Id: EXO-12-051.tex 253096 2014-07-26 20:12:19Z alverson $
\newlength\cmsFigWidth
\ifthenelse{\boolean{cms@external}}{\setlength\cmsFigWidth{0.98\columnwidth}}{\setlength\cmsFigWidth{0.6\textwidth}}
\ifthenelse{\boolean{cms@external}}{\providecommand{\cmsLeft}{top}}{\providecommand{\cmsLeft}{left}}
\ifthenelse{\boolean{cms@external}}{\providecommand{\cmsRight}{bottom}}{\providecommand{\cmsRight}{right}}
\newcommand{\CLs}{CL\ensuremath{_\mathrm{s}}\xspace}
\cmsNoteHeader{EXO-12-051}
\title{Search for jet extinction in the inclusive \texorpdfstring{jet-\pt spectrum from proton-proton collisions at $\sqrt{s}=8$\TeV}{jet-pt spectrum from proton-proton collisions at sqrt(s) = 8 TeV}}

\date{\today}

\abstract{
The first search at the LHC for the extinction of QCD jet production is presented, using data collected with the CMS detector corresponding
to an integrated luminosity of 10.7\fbinv of proton-proton collisions at a center-of-mass energy of 8\TeV.  The
extinction model studied in this analysis
is motivated by the search for signatures of strong gravity at the \TeV scale (terascale gravity) and assumes the existence of
string couplings in the
strong-coupling limit.  In this limit, the string model predicts the suppression of all high-transverse-momentum standard model
processes, including jet production, beyond a certain energy scale.  To test this prediction, the measured transverse-momentum spectrum is compared to the theoretical prediction of the standard model.  No significant deficit of events is found at high transverse momentum.   A 95\% confidence level lower
limit of 3.3\TeV is set on the extinction mass scale.
}

\hypersetup{%
pdfauthor={CMS Collaboration},%
pdftitle={Search for jet extinction in the inclusive jet-pt spectrum at sqrt(s)=8 TeV},%
pdfsubject={CMS},%
pdfkeywords={CMS, physics, jets}}

\maketitle

\section{Introduction}

The scattering of high-energy particles in theories of quantum gravity is fundamentally different from that expected by the local quantum field theories of the standard model (SM)~\cite{1999Banks}.  The Planck scale, the threshold at which quantum gravity becomes strong, is therefore a fundamental boundary beyond which some modification to the SM is required.
The Planck scale differs from the electroweak scale by 16 orders of magnitude, creating what is commonly known as the hierarchy
problem.  There are many models that propose a mechanism by which these two scales are related to one another through the
hypothesized existence of extra spatial dimensions.  Propagation of gravitons through these extra dimensions could explain the
relative weakness of gravity compared to the strong and electroweak interactions.  Depending on the model, a variety of striking signatures
of physics beyond the SM may be observed.  As a result, models that predict terascale gravity have been
the
subject of numerous searches at the CERN LHC~\cite{Black_Holes_CMS2012, Black_Holes2_CMS2012, Diphotons_CMS2012, Dijet_CMS2011, Dilepton_CMS2011, StrongGravity_ATLAS2012,Diphoton_ATLAS2012,ZZ_ATLAS2012, CMS_monojets, CMS_monophotons}.  Some of these
searches are designed to look for effects such as resonant production and decay of new
states, e.g. Randall--Sundrum gravitons~\cite{Randall:1999ee}, as well as for continuum enhancements to
SM processes
from both virtual and direct graviton production~\cite{ref:ADD}.  Direct searches for production of microscopic black holes consider events with high transverse momentum (\pt) and multiple
objects from the decay of
possible high-entropy intermediate states~\cite{1999Banks, ref:Giddings_and_Thomas,ref:Landsberg_Dimopoulos}.

As of yet, no signal indicative of terascale gravity has been found. Nevertheless, it has been suggested that evidence of terascale gravity could also be found through
more subtle effects on the jet-\pt spectrum manifesting themselves as a deviation from the predictions of quantum chromodynamics
(QCD)~\cite{1999Banks, ref:Giddings_and_Thomas, Meade:2007, Kilic:2012wp}.
While the
production of black holes or particles indicative of non-perturbative quantum gravity can have a rapidly
increasing total cross section beyond some energy scale, their decay to
isolated jets or other low-multiplicity final states
could be suppressed, leading to a full suppression of high-\pt SM scattering processes (jet extinction).
Because jet production is the leading SM process at high \pt, such effects would be initially noticeable as a jet
extinction signature~\cite{Kilic:2012wp}.
In this sense, the search for jet extinction is complementary to searches for black holes in high-multiplicity final states.  These
final states arise in the asymptotic limit, where black holes are expected to behave classically~\cite{ref:Landsberg_Dimopoulos}.  The
extinction search explores an intermediate regime, where a high-multiplicity signature may not be readily observable.

There are several models that include extinction phenomena~\cite{Meade:2007, Kilic:2012wp}.
In this, the first search for extinction effects at the LHC, we consider a model with a large-width Veneziano form factor modification of QCD processes with an
extinction mass scale $M$ equivalent to the modified Planck scale~\cite{Kilic:2012wp}.  This form factor is discussed in greater detail in
Section 3.
Beyond the scale $M$, the predominance of intermediate high-entropy string states will suppress
high-\pt SM jet
production.  This search exploits techniques developed for the measurement of the differential jet production cross section as a function of \pt at the CMS~\cite{CMS:2011ab} experiment to search for a modification of the jet-\pt spectrum consistent with extinction phenomena, in which there
are fewer
high-\pt jets than expected from the SM.   This analysis is especially sensitive to the correlations of the systematic uncertainties between bins in jet-\pt, so a detailed evaluation of the systematic
uncertainties associated with the jet energy scale (JES) and the parton distribution functions (PDFs) is performed.

\section{The CMS detector}

The central feature of the CMS detector~\cite{CMS} is a superconducting
solenoid of 6\unit{m} internal diameter, providing a field of 3.8~T. Within the field volume are silicon pixel and
strip trackers, a lead-tungstate crystal electromagnetic calorimeter (ECAL), and a brass and scintillator hadron calorimeter (HCAL).
Muons are measured in gas-ionization detectors embedded in the steel flux-return yoke.

The CMS experiment uses a right-handed coordinate system, with the origin at the nominal interaction point, the $x$ axis pointing
to the center of the LHC ring, the $y$ axis pointing up (perpendicular to the plane of the LHC ring), and the $z$ axis along the
counterclockwise-beam direction. The polar angle, $\theta$, is measured from the positive $z$ axis and the azimuthal
angle, $\phi$, is measured in the $xy$ plane.  The pseudorapidity is defined as $\eta = -\ln[\tan(\theta/2)]$.

The first level of the CMS trigger system is composed of customized hardware and uses information from the
calorimeters and muon detectors to select events of interest within a 4\unit{$\mu$s} interval following each beam crossing.
The high-level trigger~\cite{ref:HLT} (HLT) processor farm further decreases the event rate from about 100\unit{kHz} to about
400\unit{Hz} before the data are recorded for analysis.

\section{Modeling of the SM and extinction hypotheses}

The SM prediction for the jet-\pt spectrum is calculated at next-to-leading order (NLO) with the \textsc{NLOJet++}
program within the \textsc{fastNLO} framework~\cite{Nagy:2001fj, Nagy:2003tz, Britzger:2012bs}. The CT10
PDF set~\cite{ref:CT10} is used in this calculation.  The renormalization and factorization scales, $\mu_{R}$
and $\mu_{F}$, are set equal to the jet-\pt.
The NLO jet spectra do not include non-perturbative (NP) effects or any modeling of the detector response.  The NP effects, which account for hadronization and
multi-parton interactions, are incorporated as corrections determined from the \PYTHIA 6.424~\cite{ref:Pythia} Monte Carlo (MC) generator.
The generator is used to simulate QCD events with and without NP effects.  The corrections are derived from the ratio of the resulting \pt spectra.  The NP correction decreases monotonically as a function of jet-\pt, from 1.03 at 592\GeV to 1.01 at 2500\GeV.  This process is repeated using the \HERWIG 2.4.2~\cite{ref:Herwig} generator.  The difference between the corrections derived from these generators is found to be negligible in the phase space of this analysis.
The corrected NLO jet spectra are convolved with a function that models the jet energy resolution (JER) in the CMS
detector~\cite{ref:JECnew}.  These smeared spectra can be compared directly to the observed spectrum.  The smeared NLO jet spectrum is
referred to as $\rd\sigma^\mathrm{QCD}/\rd p_\mathrm{T, NLO}$.  This procedure is repeated to produce a smeared leading-order (LO) jet-\pt spectrum, labeled as $\rd\sigma^\mathrm{QCD}/\rd p_\mathrm{T, LO}$.  The predicted spectrum does not include weak radiative corrections~\cite{Dittmaier:2012kx}, but the impact of these corrections on our sensitivity to an extinction signature is evaluated during the limit-setting procedure.

The effects of extinction at LO are also modeled using the \PYTHIA MC generator.  The
matrix elements of each color channel are modified by Veneziano-type form factors~\cite{Kilic:2012wp,String_Hunters},
which affect all $2\to2$ scattering amplitudes.
The input parameters for these form factors are the extinction mass scale $M$ and
a dimensionless width parameter related to the strength of the string coupling.  For small values of the width parameter, these form
factors are similar to those that describe
string resonances~\cite{Mirabelli:1998rt,String_Hunters}.  This is referred to as the weak-coupling limit.
The regime where the width parameter is close to unity is known as the strong-coupling limit.  In this limit, extinction physics rapidly overwhelms LO SM processes as well as any resonant string production.
Beyond the scale $M$, scattering processes are dominated by a continuum of high-entropy intermediate states,
which results in suppression of SM jet production~\cite{Kilic:2012wp}.  This search assumes a width parameter of one, the
absolute strong-coupling limit of the string model.  Values of the width above one represent a very different phenomenology where the form factors no longer monotonically decrease as a function of jet momentum.  This range of the width parameter has not been studied in this analysis.

The effects of extinction are predominantly found in $2\to2$ scattering processes.  Such processes are dominated by the LO calculation at a given \pt scale.  The signal is approximated with a LO generator.  The extinction process is assumed to have a very weak effect on higher-order interactions.
A sigmoid function provides a good functional fit of the effect of the Veneziano form factors on the LO jet-\pt
spectrum~\cite{Kilic:2012wp}:

\begin{linenomath}
\begin{equation}
F(\pt, M) = \frac{1}{1+\exp{\frac{\pt - p_{\mathrm{T}, 1/2}(M)}{p_{\mathrm{T},0}(M)}}}.
\end{equation}
\end{linenomath}

Here, $p_{\mathrm{T},1/2}$ describes the \pt threshold at which LO jet production is reduced to half the SM expectation, while $p_{\mathrm{T},0}$
indicates how quickly the LO cross section exponentially falls relative to the SM prediction.  
This relation yields the following equation for the jet-\pt spectrum assuming extinction at LO, where $\sigma^{\rm Ext}$ is the jet production
cross section assuming extinction:

\begin{linenomath}
\begin{equation}
\frac{\rd\sigma^\text{Ext}}{\rd p_\mathrm{T, LO}} = \frac{\rd\sigma^\mathrm{QCD}}{\rd p_\mathrm{T, LO}} F(\pt, M)
\end{equation}
\end{linenomath}

and at NLO:

\begin{linenomath}
\begin{equation}
\frac{\rd\sigma^\text{Ext}}{\rd p_\mathrm{T, NLO}} = \frac{\rd\sigma^\mathrm{QCD}}{\rd p_\mathrm{T, NLO}} - \frac{\rd\sigma^\mathrm{QCD}}{\rd p_\mathrm{T, LO}} + \frac{\rd\sigma^\text{Ext}}{\rd p_\mathrm{T, LO}}.
\end{equation}
\end{linenomath}

Several simulations of LO jet production are performed, assuming values of $M$ between 2 and 5\TeV in increments of 500\GeV.  The jet-\pt
spectrum is produced at NLO for each sample using NP corrections and resolution smearing as described above.  The values of $p_{\mathrm{T},1/2}(M)$ and $p_{\mathrm{T},0}(M)$ are extracted from a fit of $F(\pt, M)$ to the expected \pt distribution for each value of $M$.  The intermediate values of $p_{\mathrm{T},1/2}(M)$ and $p_{\mathrm{T},0}(M)$ are interpolated between these fitted points.  The fitted value of $p_{\mathrm{T},0}(M)$ is nearly independent of $M$ and ranges between 260 and 330 \GeV, while $p_{\mathrm{T},1/2}(M)$ is about half of $M$.  The systematic uncertainty associated with the choice of fit is negligible.

For finite values of
$M$, the predicted jet-\pt spectrum is suppressed in systems with an invariant mass above $M$.  At very large values of $M$, the SM and
extinction spectra become identical.

\section{Event reconstruction and selection}

A particle-flow algorithm~\cite{ref:PFlow2, ref:PFlow} is used to reconstruct the events.  Jets are formed by clustering the reconstructed
particle-flow objects using the anti-\kt algorithm~\cite{Cacciari:2008gp} with a distance parameter $R$ of 0.7.  This value is larger
than the usual distance parameter of 0.5 used in most CMS analyses.  The larger cluster size reduces the likelihood that jets will be lost
because of detector effects.
The jet transverse momentum resolution is typically 15\% at \pt = 10\GeV, 8\% at 100\GeV, and 4\% at 1\TeV. Jet energy
corrections are
derived from simulation and are confirmed with measurements of energy balance in recorded dijet and photon+jet events. The combined corrections are approximately 5--10\%, depending on the pseudorapidity and \pt of the jet.  To suppress spurious signals from detector
noise~\cite{ref:CMSNoise}, jets are required to satisfy stringent selection criteria~\cite{ref:jetid}.  Specifically, each jet must contain at least two particles, one of
which is a charged
hadron.  Additionally, each of the jet energy fractions carried by neutral hadrons, photons, electrons, and muons must be less than 90\%.  This analysis is conducted in a regime where the purity and acceptance of the jets in data are both close to unity, and therefore no systematic uncertainty is attributed to the selection criteria.

The data used in this analysis were collected from an HLT trigger that accepted events containing at least one
jet with $\pt>320$\GeV.  An offset is applied to trigger-selected jets to subtract the energy deposited as a result of additional interactions per beam
crossing (pileup); this offset does not affect the trigger efficiency.
Events with objects originating from an interaction within an LHC beam crossing are selected by requiring the presence of at least one
primary vertex within 24\unit{cm} of the detector center along the $z$ axis.  The primary event vertex is chosen from all reconstructed vertices
by
selecting the one with the largest sum of the $\pt^{2}$ of all associated tracks.  For the purpose of additional noise suppression, the missing
transverse energy,
defined as the
magnitude of the vector sum \pt of all reconstructed
particle-flow objects, must be less than 30\% of the total transverse energy deposited in the detector.  All jets in each event that pass the selection criteria are binned as a function of jet-\pt, following a convention adopted by other inclusive-jet analyses in CMS.  The bin widths are
variable, increasing with jet-\pt and
corresponding approximately to the jet-\pt resolution~\cite{CMS:2011ab}. Jets are required to have $\pt>592$\GeV and pseudorapidity
$\abs{\eta}<1.5$ to
ensure that the
trigger is at least 99\% efficient in all \pt bins used.  This search is performed in 18 \pt bins between 592 and 2500 \GeV.

\begin{figure}[htp]
\centering
\includegraphics[width=\cmsFigWidth]{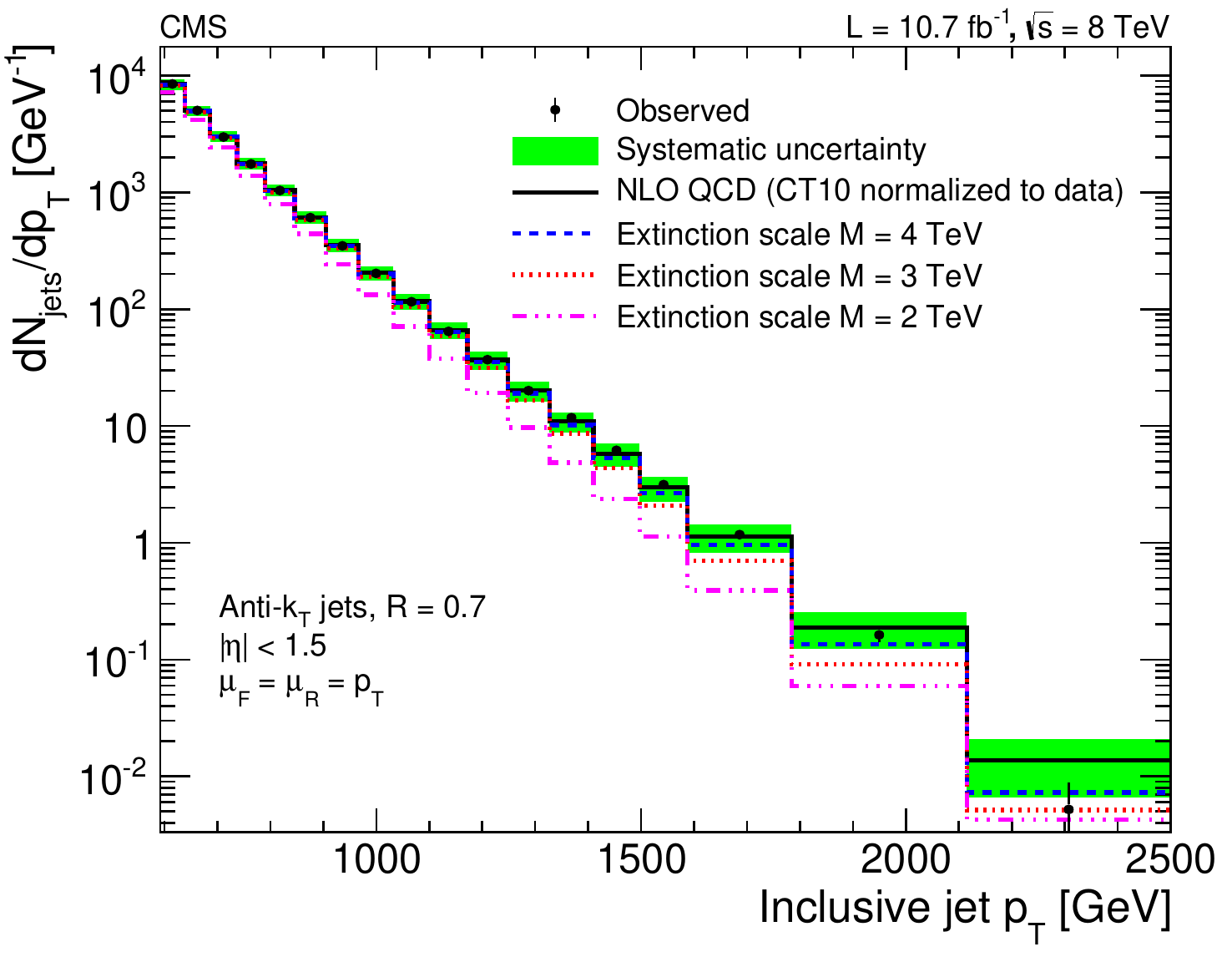}
\caption{Inclusive jet-\pt spectrum (points) for $\abs{\eta} < 1.5$, as observed in data.   The SM NLO
simulation with non-perturbative corrections, convolved with the detector response and
normalized to the total number of jets observed in data, is shown by the solid line.  The spectra predicted by the extinction model are defined relative to the SM prediction as described by Eq. 3 for the values of $M = 2, 3,$ and $4\TeV$ and shown by the dashed lines.  The colored band shows the magnitude of the sources of
systematic uncertainty added in quadrature.  These sources include the JES, JER, PDFs, and scale variations.  An additional source of systematic uncertainty is attributed to the integrated luminosity during all formal comparisons between the data and models, but has little impact on the sensitivity to an extinction signature.
The renormalization scale ($\mu_\mathrm{R}$) and factorization scale ($\mu_\mathrm{F}$) are set to the \pt of the hard-scattered parton.}
\label{fig:inclusive_jet_pt}

\end{figure}

\begin{figure}[htp]
\centering
\includegraphics[width=\cmsFigWidth]{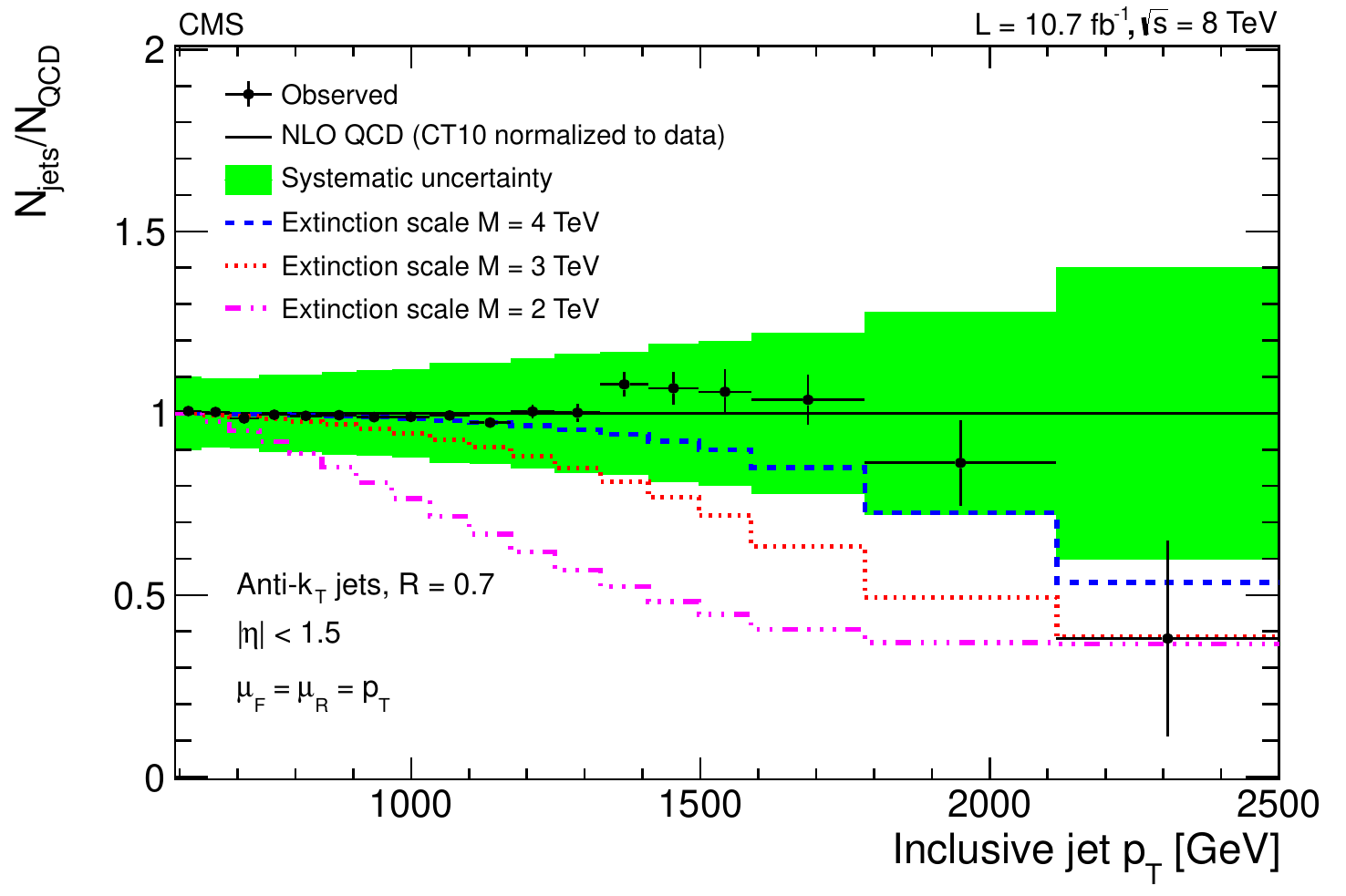}
\caption{The ratio of the inclusive jet $\pt$ spectrum to the NLO QCD prediction with non-perturbative corrections and convolved with the
detector resolution.  The horizontal bars on the data indicate the width of each bin in \pt.  The colored band shows the quadratic sum of the sources of systematic uncertainty,
including JES, JER, PDFs, and scale variations.  The uncertainty in the integrated luminosity is excluded, as the model predictions have been normalized to the number of jets observed in data.  The dashed lines indicate the effects of extinction at three different values of the extinction mass scale, $M=2$, 3, and 4\TeV.
}\label{fig:jet_pt_ratios}
\end{figure}

A comparison between the observed inclusive jet-\pt spectrum and the spectrum predicted at NLO with the CT10 PDF set is shown in
Figs.~\ref{fig:inclusive_jet_pt} and~\ref{fig:jet_pt_ratios}.  The predicted spectrum includes non-perturbative corrections and smearing by the detector response, and is normalized to the total number
of jets in data that pass all selection criteria.  However, in the comparison of the model to the data as described in Section 5, the SM distribution is instead normalized to the number of jets expected given an integrated luminosity of 10.7\fbinv.  The number of jets observed in data is 3\% lower than the number expected assuming the CT10 PDF set at NLO.  This discrepancy is attributed to uncertainty in the PDF parameters, scale variations in the cross section calculation, or uncertainty in the total integrated luminosity.  As the search for an extinction signature is only concerned with the shape of the jet-\pt spectrum, a small shift in the absolute normalization has little impact on the sensitivity.  In Figs. 1 and 2 the data and the extinction model are compared after any differences in the normalization have been resolved.  In these figures, the quadratic sum of all sources of systematic uncertainty is shown.  The total systematic uncertainty includes contributions from both theoretical and experimental sources.
The theoretical uncertainty is composed of the uncertainty from the PDFs as well as the uncertainty obtained by varying the renormalization and factorization scales. The experimental uncertainty is derived from
the uncertainties in the JES and JER.  During the formal comparison of the model to data where the predicted spectrum is not normalized to the number of jets observed, an additional source of uncertainty is attributed to the integrated luminosity.  Figure~\ref{fig:jet_pt_ratios} shows the ratio of the inclusive spectrum to the SM NLO expectation and includes the predicted spectra from the extinction model for three different values of the extinction mass scale $M$.  

\section{Statistical method and systematic uncertainties}

To distinguish between SM NLO jet production and the alternative hypothesis (jet extinction), a profile-likelihood
ratio test statistic~\cite{TestStatLHC} is constructed as a function of a signal strength parameter, $\beta\equiv M^{-2}$.
The variable $\beta$ is chosen so that as $\beta\to0$ the extinction model approaches the SM prediction.

We set limits using the modified-frequentist criterion \CLs~\cite{Junk:1999kv, LEP-CLs}.
All sources of systematic uncertainty are treated as nuisance parameters with log-normal prior constraints and are constructed in the likelihood
to have the same value across all jet-\pt bins.  This construction implicitly assumes that the systematic uncertainties are completely
correlated in jet
\pt.

To account for correlations in the JES and PDF uncertainties between \pt bins, the uncertainties are subdivided into their underlying components.  These individual components are strongly correlated across
all \pt bins and tend to be dominant at different values of jet-\pt.  As an example, uncertainties in the gluon PDF will be dominant at low
\pt compared to uncertainties in the quark PDFs.  The JES uncertainty is decomposed into each of its orthogonal sources.  For the PDF uncertainty, the contributions from each of
the eigenvectors in the CT10~\cite{ref:CT10} PDF set are evaluated separately.  As a crosscheck, the search is repeated with respect to the
MSTW2008~\cite{ref:MSTW2008} PDF
set.  Among the PDF sets in common use, the CT10 set predicts the highest inclusive jet cross section at high \pt, while the MSTW2008 set gives one of the lowest.  The
results derived with respect to these two PDF sets serve as bounds on the result expected when using other sets, including those which are used in comparison to dedicated measurements of the inclusive jet production cross section~\cite{CMS:2011ab}, such as NNPDF~\cite{ref:NNPDF2}, HERA~\cite{ref:HERA}, or ABKM~\cite{ref:ABKM}.

The CT10 PDF set comprises a central prediction and 26 eigenvectors.  The central prediction assumes all PDF input parameters are set to
their central values.
Each eigenvector pair corresponds to the upward and downward uncertainty in one of those input parameters.  The difference between the
predictions of each eigenvector pair and the central prediction is taken as a source of systematic uncertainty at ${\pm}1\sigma$.  A
source of systematic uncertainty is defined as non-trivial if, at one standard deviation in either direction, it produces a shift in any \pt bin greater
than 1\% of the occupancy given by the central prediction.  Under this definition, 15 of the 26 CT10 eigenvectors are found to be non-trivial.

\begin{figure}[htp]
\centering
\includegraphics[width=\cmsFigWidth]{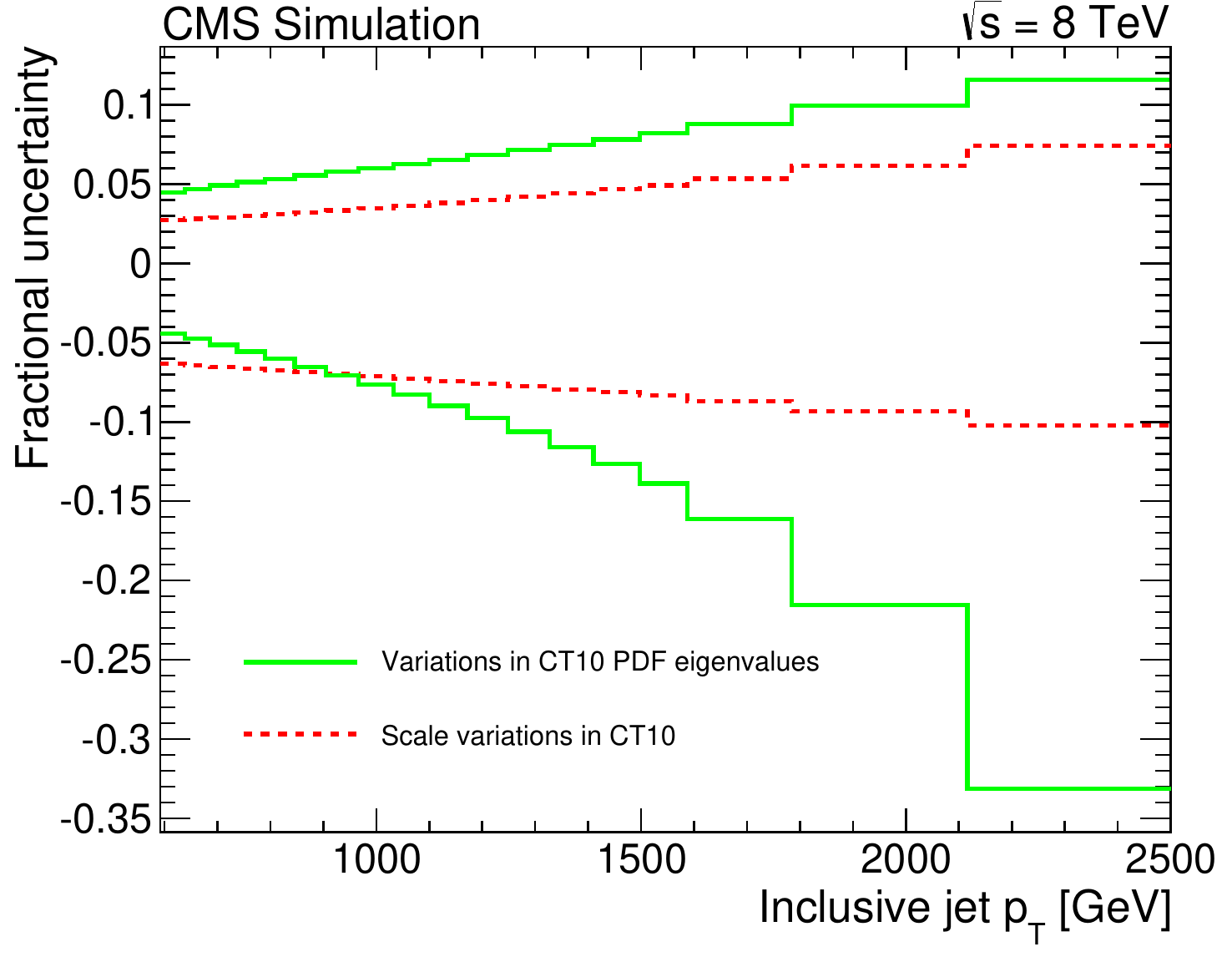}

\caption{Uncertainty at ${\pm}1$ standard deviation described by the combined variations of all CT10 PDF eigenvectors added in quadrature (solid lines),
as well as the scale variations (dotted lines).  The
uncertainty is expressed as a fraction of the central occupancy of each \pt bin.  For the fit of the model to data and the limit setting procedure, the PDF uncertainty is subdivided into individual sources for each eigenvector pair.
}\label{fig:sys_theory}
\end{figure}

The relative uncertainty described by the combined variation of these eigenvector sets in quadrature and the scale variations are shown in
Fig.~\ref{fig:sys_theory} as a function of jet-\pt.
The uncertainties associated with the renormalization and factorization scales are computed by varying the scales coherently up and
down by a factor of 2.
As the effect of extinction on the jet-\pt spectrum is expressed relative to the SM prediction, by construction the PDF variations do not
affect any of the extinction parameters.

\begin{figure}[htp]
\centering
\includegraphics[width=\cmsFigWidth]{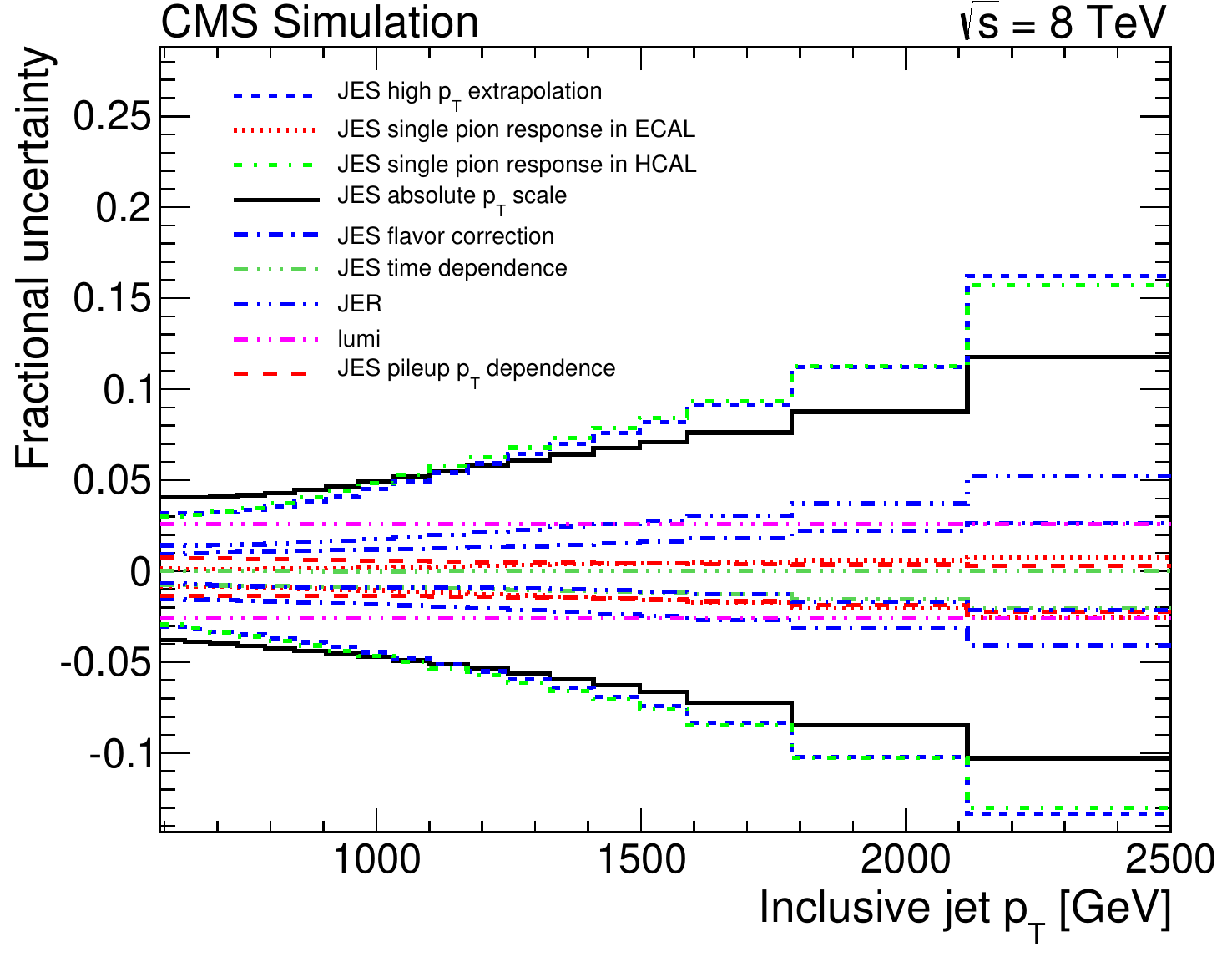}
\caption{Systematic uncertainty from all experimental sources at ${\pm}1$ standard deviation, expressed as a fraction of the central occupancy of
each \pt bin.  The luminosity uncertainty is constant in jet-\pt, while the JES and JER uncertainties are modelled as transfer matrices between all \pt
bins.  The seven non-trivial sources of JES uncertainty are shown (out of 19 total).
}\label{fig:sys_exp}
\end{figure}

Given the exponentially falling nature of the inclusive jet-\pt spectrum, the JES is one of the dominant
sources of systematic uncertainty.  The JES uncertainty is composed of 19 orthogonal sources.  Of these, seven are found
to be non-trivial according to the criterion defined above:
the absolute \pt scale; the single pion response in the ECAL; the single pion response in the HCAL; the flavor composition correction; the
time dependence;
the pileup \pt scale; and the extrapolation of the absolute scale into the high-\pt regime~\cite{ref:JECnew}.
The effects of JER are also included as nuisance parameters.
The uncertainty in luminosity is taken as a constant scale factor with a 2.6\% relative uncertainty~\cite{CMS-PAS-SMP-12-008}.  The
relative uncertainty of all non-trivial detector-related sources of systematic uncertainty (JES, JER, and integrated luminosity) is shown
in Fig.~\ref{fig:sys_exp} as a function of jet-\pt.

Including systematic
uncertainties, the best-fit value of $\beta$ is $(0.008\pm0.033)\TeV^{-2}$,
which is consistent with the SM expectation.

\begin{figure}[htp]
\centering
\includegraphics[width=\cmsFigWidth]{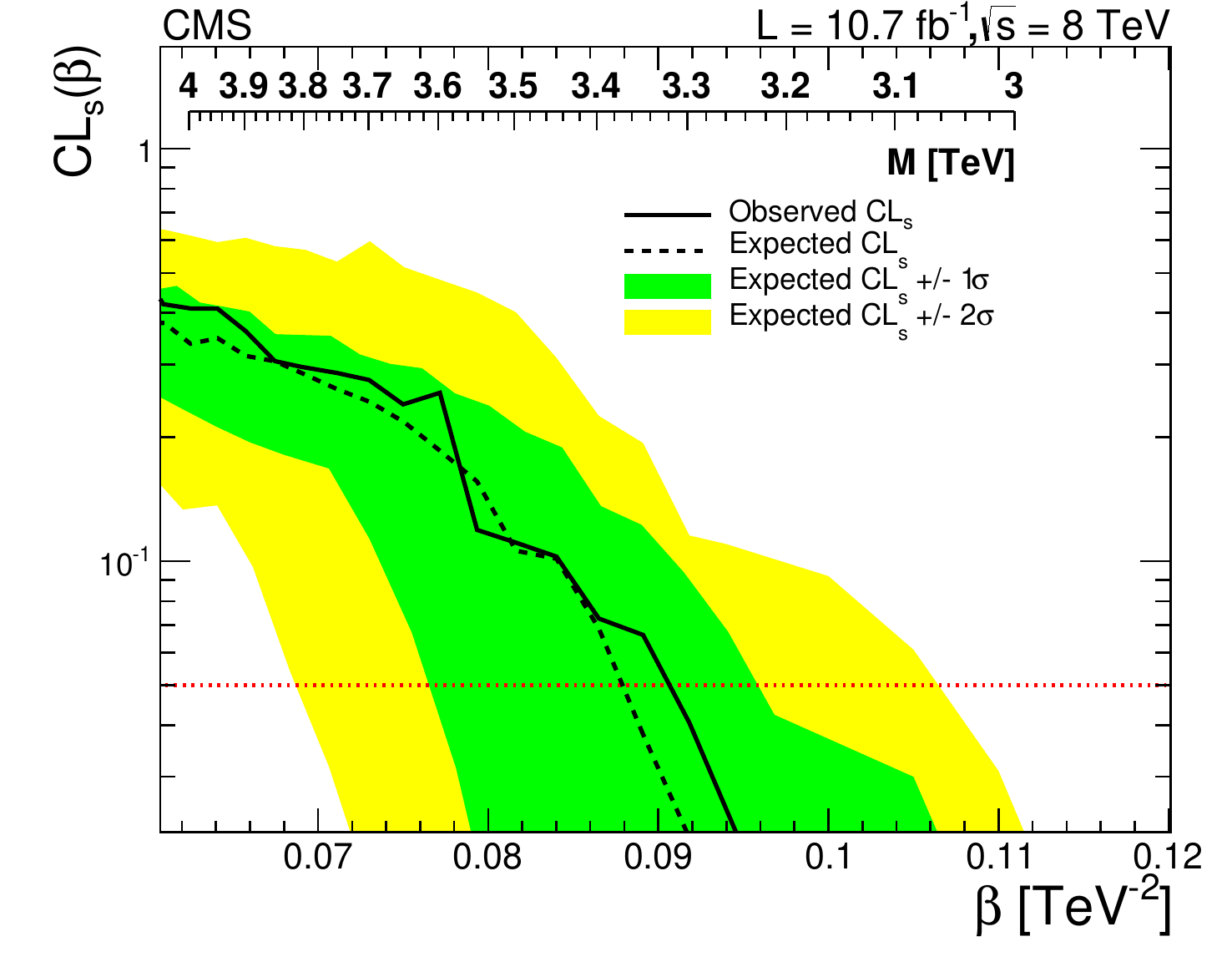}
\caption{The results of a \CLs scan in the extinction mass scale,
$\beta = M^{-2}$.
The observed dependence of \CLs on $\beta$ is shown by the solid line.  The observed upper limit on $\beta$ is $0.090\TeV^{-2}$ at 95\% CL (indicated by the horizontal dotted line), corresponding to a lower limit of 3.3\TeV on the extinction mass scale $M$.  The dashed line indicates the expected median of results for the SM
hypothesis, while the green (dark) and yellow (light) bands indicate the quantiles, which contain 68\% and 95\% of the expected results, respectively.
}\label{fig:CLSLimit}
\end{figure}

The dependence of \CLs on the parameter $\beta$ is shown in
Fig.~\ref{fig:CLSLimit}.  The observed upper limit on $\beta$ is $0.090\TeV^{-2}$  at 95\% confidence level (CL), translating to an observed lower limit on $M$ of 3.3\TeV.  The expected upper limit on $\beta$ is $0.088\TeV^{-2}$  at 95\% CL, corresponding to an expected lower limit on $M$ of 3.4\TeV.  These relatively close expected and observed values reflect good agreement between the observed data and the null hypothesis.

As an additional check, the limit setting procedure is repeated using the MSTW2008 PDF set~\cite{ref:MSTW2008} to derive the SM hypothesis.  The limits obtained using the CT10 and MSTW2008 PDFs agree to within 10\%.  As the MSTW2008 PDFs predict a
lower cross
section at
very high jet-\pt compared to CT10, the limit produced in this check is less conservative.

Finally, the limits have been calculated including weak radiative corrections to the SM prediction, with a decrease of less than 100 GeV to the exclusion region.

\section{Summary}

The first search for the extinction of jet production has been performed at the LHC using proton-proton collision data at $\sqrt{s}=8\TeV$
collected by the CMS detector and corresponding to an integrated luminosity of 10.7\fbinv.
The extinction model studied in this analysis is motivated by the search for signatures of terascale gravity
at the LHC and assumes the existence of string couplings in the strong-coupling limit.  In this limit, the string model predicts suppression of high-\pt jet production beyond an extinction mass scale $M$.
A detailed comparison
between the measured \pt spectrum and the theoretical prediction is conducted.  No
significant deficit of events is found at high transverse momentum.  A 95\% confidence level lower limit of 3.3\TeV is set on the extinction mass scale $M$.

\section*{Acknowledgements}

We congratulate our colleagues in the CERN accelerator departments for the excellent performance of the LHC and thank the technical and administrative staffs at CERN and at other CMS institutes for their contributions to the success of the CMS effort. In addition, we gratefully acknowledge the computing centres and personnel of the Worldwide LHC Computing Grid for delivering so effectively the computing infrastructure essential to our analyses. Finally, we acknowledge the enduring support for the construction and operation of the LHC and the CMS detector provided by the following funding agencies: BMWFW and FWF (Austria); FNRS and FWO (Belgium); CNPq, CAPES, FAPERJ, and FAPESP (Brazil); MES (Bulgaria); CERN; CAS, MoST, and NSFC (China); COLCIENCIAS (Colombia); MSES and CSF (Croatia); RPF (Cyprus); MoER, ERC IUT and ERDF (Estonia); Academy of Finland, MEC, and HIP (Finland); CEA and CNRS/IN2P3 (France); BMBF, DFG, and HGF (Germany); GSRT (Greece); OTKA and NIH (Hungary); DAE and DST (India); IPM (Iran); SFI (Ireland); INFN (Italy); NRF and WCU (Republic of Korea); LAS (Lithuania); MOE and UM (Malaysia); CINVESTAV, CONACYT, SEP, and UASLP-FAI (Mexico); MBIE (New Zealand); PAEC (Pakistan); MSHE and NSC (Poland); FCT (Portugal); JINR (Dubna); MON, RosAtom, RAS and RFBR (Russia); MESTD (Serbia); SEIDI and CPAN (Spain); Swiss Funding Agencies (Switzerland); MST (Taipei); ThEPCenter, IPST, STAR and NSTDA (Thailand); TUBITAK and TAEK (Turkey); NASU and SFFR (Ukraine); STFC (United Kingdom); DOE and NSF (USA).

Individuals have received support from the Marie-Curie programme and the European Research Council and EPLANET (European Union); the Leventis Foundation; the A. P. Sloan Foundation; the Alexander von Humboldt Foundation; the Belgian Federal Science Policy Office; the Fonds pour la Formation \`a la Recherche dans l'Industrie et dans l'Agriculture (FRIA-Belgium); the Agentschap voor Innovatie door Wetenschap en Technologie (IWT-Belgium); the Ministry of Education, Youth and Sports (MEYS) of the Czech Republic; the Council of Science and Industrial Research, India; the HOMING PLUS programme of Foundation for Polish Science, cofinanced from European Union, Regional Development Fund; the Compagnia di San Paolo (Torino); and the Thalis and Aristeia programmes cofinanced by EU-ESF and the Greek NSRF.
\bibliography{auto_generated}   

\cleardoublepage \appendix\section{The CMS Collaboration \label{app:collab}}\begin{sloppypar}\hyphenpenalty=5000\widowpenalty=500\clubpenalty=5000\textbf{Yerevan Physics Institute,  Yerevan,  Armenia}\\*[0pt]
V.~Khachatryan, A.M.~Sirunyan, A.~Tumasyan
\vskip\cmsinstskip
\textbf{Institut f\"{u}r Hochenergiephysik der OeAW,  Wien,  Austria}\\*[0pt]
W.~Adam, T.~Bergauer, M.~Dragicevic, J.~Er\"{o}, C.~Fabjan\cmsAuthorMark{1}, M.~Friedl, R.~Fr\"{u}hwirth\cmsAuthorMark{1}, V.M.~Ghete, C.~Hartl, N.~H\"{o}rmann, J.~Hrubec, M.~Jeitler\cmsAuthorMark{1}, W.~Kiesenhofer, V.~Kn\"{u}nz, M.~Krammer\cmsAuthorMark{1}, I.~Kr\"{a}tschmer, D.~Liko, I.~Mikulec, D.~Rabady\cmsAuthorMark{2}, B.~Rahbaran, H.~Rohringer, R.~Sch\"{o}fbeck, J.~Strauss, A.~Taurok, W.~Treberer-Treberspurg, W.~Waltenberger, C.-E.~Wulz\cmsAuthorMark{1}
% ...
\vskip\cmsinstskip
% ...
1:~~Also at Vienna University of Technology, Vienna, Austria\\
2:~~Also at CERN, European Organization for Nuclear Research, Geneva, Switzerland\\
% ...

\end{sloppypar}
\end{document}